\documentclass{endm}
\usepackage{endmmacro}
\usepackage{graphicx}
\usepackage{subfigure}

\usepackage{t1enc}

\graphicspath{{figures/}}

\begin{document}

% DO NOT REMOVE: Creates space for Elsevier logo, ScienceDirect logo
% and ENDM logo
\begin{verbatim}\end{verbatim}\vspace{2.5cm}

\begin{frontmatter}

\title{On some simplicial elimination schemes for chordal graphs}

\author{Michel Habib \thanksref {myemail}}
\address{LIAFA, CNRS and Universit\'e Paris Diderot - Paris 7, France}

\author{Vincent Limouzy \thanksref{coemail}}
\address{Dept. of Computer Science, University of Toronto, Canada}

\thanks[myemail]{Email:
   \href{mailto:habib@liafa.jussieu.fr} {\texttt{\normalshape
  habib@liafa.jussieu.fr }}} \thanks[coemail]{Email:
   \href{mailto:limouzy@cs.toronto.edu} {\texttt{\normalshape
   limouzy@cs.toronto.edu}}}

\begin{abstract}
We introduced here an interesting tool for the structural study of chordal graphs, namely the \textbf{Reduced Clique Graph}. Using some of its  combinatorial properties we 
show  that  for any chordal graph we can construct in linear time a simplicial  elimination scheme starting with a pending maximal clique attached via  a minimal separator  maximal under inclusion among all minimal separators.
\end{abstract}


\end{frontmatter}

\section{Introduction}\label{intro}
In the following text, a graph is always finite, simple, loopless, undirected and connected. A graph is \textbf{chordal} iff it has no chordless cycle of length $\geq 4$.
The class of chordal graphs is one of the first class to have been studied at the beginning of the theory of perfect graphs. Since then chordal graphs have been intensively studied, as can be seen in the following books \cite{Golumbic80,BLS99}.

Let us recall the main notions defined for chordal graphs. A \textbf{maximal clique} of $G$ is a complete subgraph maximal under inclusion.
A \textbf{minimal separator} is  a subset of vertices $S$ 
for which  it exist $a, b \in G$ such that $a$ and $b$ are not connected in $G-S$,
and $S$ is minimal under inclusion with this property.
A vertex is \textbf{simplicial} if its neighborhood is a clique (complete graph).
An ordering $x_{1}, \dots, x_{n}$ of the vertices is a \textbf{simplicial elimination scheme}, if for every $i \in [1,n-1]$
$x_{i}$ is a simplicial vertex in $G[x_{i+1}, \dots x_{n}]$.
A \textbf{maximal clique tree}   is a tree $T$ that satisfies
the following three conditions:
Vertices of $T$ are associated with the maximal cliques of $G$.
Edges of $T$ correspond to minimal separators.
For any vertex $x \in G$, the cliques containing $x$ yield a subtree of $T$.

Using results of Dirac \cite{D61}, Fulkerson, Gross \cite{FG65}, Buneman \cite{B74}, Gavril \cite{G74} and Rose, Tarjan and Lueker \cite{RTL76},
we have:
\begin{theorem} The following  5 statements are equivalent and characterize chordal graphs. 
\begin{description}
\item[(i)]
G  has a simplicial elimination scheme
\item[(ii)]
Every minimal separator is a clique
\item[(iii)]
$G$ admits a maximal clique tree.
\item[(iv)] $G$ is the intersection graph of subtrees in a tree.
\item[(v)]
Any LexBFS provides a simplicial elimination scheme.
\end{description}

\end{theorem}
\section{The Reduced Clique Graph}
\begin{definition}
For a chordal graph $G$, we denote by $\mathcal{C}$ the set of maximal cliques of  $G$ and by $\mathcal{C}_{r}(G)$ the  \textbf{reduced clique graph}, i.e. the graph whose vertices are the maximal cliques of $G$, and two cliques are joined by an edge iff their intersection  separates them (i.e.  if  for every $x \in C- (C\cap C')$ and every $y \in C'-(C\cap C')$, $C\cap C'$ is a minimal separators for $x$ and $y$ in $G$). \end{definition}
Clearly $\mathcal{C}_{r}(G)$ is a subgraph of the intersection graph of the maximal cliques of $G$.
Each edge  $CC'$ of $\mathcal{C}_{r}(G)$ can be  labelled with the minimal separator $S=C\cap C'$. 

\begin{figure}[h!]
	\begin{center}
		\subfigure[][]{\label{fig:ex-a}\includegraphics[scale=1.3]{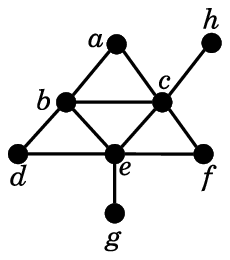}}
		\quad\quad
		\subfigure[][]{\label{fig:ex-b}\includegraphics[scale=1.3]{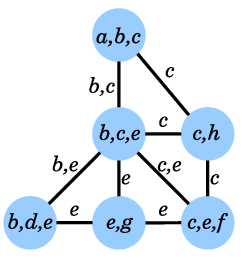}}
		\quad\quad
		\subfigure[][]{\label{fig:ex-c}\includegraphics[scale=1.3]{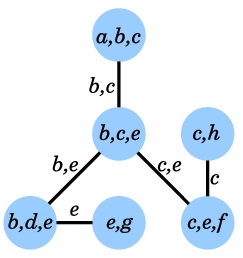}}
		\quad\quad
		\subfigure[][]{\label{fig:ex-d}\includegraphics[scale=1.3]{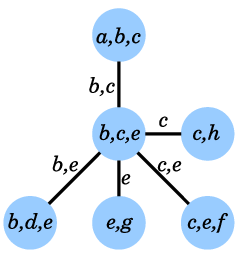}}
	\caption{An example of a chordal graph \subref{fig:ex-a}, its reduced clique-graph \subref{fig:ex-b}, note that although the maximal cliques $\{b, d, e\}$ and $\{c, e, f\}$ intersect the corresponding edge is missing. Two maximal clique-trees are shown \subref{fig:ex-c}-\subref{fig:ex-d}.}
	\label{fig:}
	\end{center}
\end{figure}

\begin{lemma}\cite{GHP95}
Let us consider three maximal cliques $C_{1}, C_{2}, C_{3}$ in  $G$,  such that  $S=C_{1}\cap C_{2}$   and $U=C_{2}\cap C_{3}$ are minimal separators in $G$, then 
 $S \subset U$ implies that  $C_{1}\cap C_{3}$ is a minimal separator of $G$.
\end{lemma}

\begin{lemma} \cite{GHP95}
Let us consider  a triangle in $\mathcal{C}_{r}(G)$ together with its 3 minimal separators labelling its edges. Then two of these minimal separators must be equal and included in the third.
\end{lemma}

With these two lemmas it is easy to prove the following result:
\begin{proposition}\cite{BP93,GHP95}
For a chordal graph $G$ maximal clique trees correspond to maximum spanning trees of $\mathcal{C}_{r}(G)$ when the edges are labelled with the size of the minimal separator they are associated with. Furthermore $\mathcal{C}_{r}(G)$ is the union of all maximal clique trees of $G$. 
\end{proposition}

As a consequence, all maximal clique trees define the same multiset of minimal separators, and from one maximal clique tree to another we can proceed by exchanging edges (with same label) on triangles.
But the graph $\mathcal{C}_{r}(G)$ has still more combinatorial properties, that we now consider.
Let us now study the limit case of the two previous lemmas, when $S=U$. First we need a basic separating  lemma (which can also be found in a more general setting of tree decompositions, see  lemma 12.3.1 in \cite{D97}).

\begin{lemma}\label{separating} Separating lemma

Let $T$ be a maximal clique tree and $C_{1}C_{2}$ and edge of $T$. Let  $T_{1}$ and $T_{2}$ the two connected components of $T-C_{1}C_{2}$. If we define $V_{i}$ for i=1,2  the union of all maximal cliques in $T_{i}$. Then $S=C_{1} \cap C_{2}$ separates every $x \in V_{1}-S$ from any $y \in V_{2}-S$.
\end{lemma}
\begin{lemma}
Let us consider three maximal cliques $C_{1}, C_{2}, C_{3}$ in  $G$,  such that  $S=C_{1}\cap C_{2}=U=C_{2}\cap C_{3}$ are minimal separators in $G$, then  either  the edge $C_{1}C_{3} \in \mathcal{C}_{r}(G)$ or  the two edges $C_{1}C_{2}, C_{2}C_{3}$ cannot belong both
to a same  maximal clique tree.
 \end{lemma}

\begin{proof} 
Suppose that the edge $C_{1}C_{3}$ does not belong to $ \mathcal{C}_{r}(G)$, i.e. that $S=C_{1}\cap C_{3}$ does not separate $C_{1}-S$ from $C_{3}-S$.  Therefore if it exists some maximal clique tree $T$  containing both edges $C_{1}C_{2}, C_{2}C_{3}$,  this would contradict the above separating lemma \ref{separating}.
\end{proof}

\begin{lemma}
Let us consider three maximal cliques $C_{1}, C_{2}, C_{3}$ in  $G$,  such that  $S=C_{1}\cap C_{2}=U=C_{2}\cap C_{3}$ are minimal separators in $G$,  if the edges  $C_{1}C_{2}, C_{2}C_{3}$  belong both to a same  maximal clique tree $T$. Then  $C_{1}C_{3} \in 
\mathcal{C}_{r}(G)$ and  $C_{1} \cap C_{3} =U$
\end{lemma}

\begin{proof} Using the previous lemma necessarily $C_{1}C_{3} \in \mathcal{C}_{r}(G)$, but lemma 1 just states that 
$C_{1} \cap C_{3} \subseteq U=S $.  If this is a strict inclusion then one can build a new maximal clique tree $T'$ by exchanging the edges 
$C_{1}C_{2}$ by $C_{1}C_{3}$. But then $T'$ would be a better spanning tree than $T$ which contradicts  the optimality of $T$ and therefore $C_{1} \cap C_{3}= U=S $.
\end{proof}

\section{Min-max separators}

For a finite chordal graph  $G$,  let us call a  min-max (resp. min-min) separator  $S$, a minimal separator that is maximal (resp. minimal) under inclusion among all minimal separators of $G$. 

\begin{theorem}\cite{LMP08}
Let $G$ be a chordal graph, then it exists a maximal clique-tree $T$ that admits a pending edge labelled with a min-max  separator.
\end{theorem}

\begin{figure}[h!]
	\begin{center}
		\includegraphics[scale=1.0]{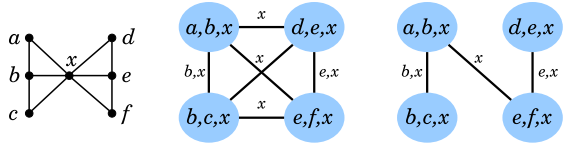}
	\caption{Preissmann's counter example \cite{Preissmann09}, A graph, its reduced clique graph and one maximal 
	clique tree}
	\label{fig:CEMinMin}
	\end{center}
\end{figure}

\begin{proof}
The proof will proceed by transforming a maximal clique tree using the above lemmas.
Let us consider   $T$ a maximal clique tree of $G$ and   some edge $ab \in T$ labelled with a min-max separator $S$.
First we need to define an operation on cliques trees, namely the \textbf{chain-reduction}. Suppose $ab$ is not a pending edge in $T$, therefore $T-\{ab\}$ is the disjoint union of two non empty trees $T_{a}, T_{b}$. If one of these trees, say $T_{a}$ admits a lead edge $xy$ labelled with a minimal separator $S' \subset S$ ($y$ being the pending clique in $T$). Then the whole chain in $T_{a}$ joining $ab$ to $xy$ is labelled with minimal separators containing $S'$. Using this fact and successive applications of the above lemmas, we can interchange in $T_{a}$ the edges $xy$ and $ay$ (or equivalently in $T$ exchanging $xy$ by $by$).
Let us go back to the proof of the theorem. If one of the subtrees $T_{a}, T_{b}$, say $T_{a}$ is made up with edges labelled with  minimal separators included in $S$, then using the chain-reduction operation we can produce another maximal clique tree $T'$ in which all the edges of $T_{a}$ are leaves attached to $b$ and $ab$ is a leaf and we have finished.
Else it exists in one of the subtrees $T_{a}, T_{b}$, say $T_{a}$, some edge $zt$ labelled with $S'$ which is not comparable with $S$.
We recurse on the maximal minimal separator that contains $S'$ and which necessarily belongs to $T_{a}$. This process necessarily ends by finding a leaf in the tree which is labelled with a max-min separator, because each time we recurse  on a strict subtree.
\end{proof}
Such maximal clique trees seem to play an important role for the study of path graphs \cite{LMP08}.
The above proof also suggests a dual result for min-min separators.
But as it was noticed by M. Preissmann \cite{Preissmann09},  such a maximal clique tree does not always 
exist. The graph depicted in figure \ref{fig:CEMinMin} does not admit a min-min elimination scheme.

\begin{figure}[h!]
	\begin{center}
		\includegraphics[scale=1.50]{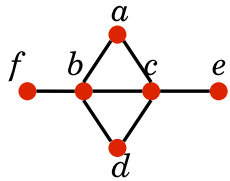}
	\caption{An exemple of graph on which $MCS$, $LexBFS$ fail to find a max-min simplicial vertex.
	For any starting vertex, both searches will end on $e$ of $f$.}
	\label{fig:MaxFail}
	\end{center}
\end{figure}

Using the above constructive proof, a polynomial scheme can be obtained to compute a min-max elimination schemes. As shown in Figure \ref{fig:MaxFail}, classical graph searches do not provide such elimination scheme.

\begin{corollary}
Such trees can be obtained in linear time.
\end{corollary}

\begin{proof}
We prove the result in the min-max case.
To obtain such a tree we can first compute a maximal clique tree $T$ of $G$ as explained in \cite{GHP95}, with its edges being labelled with the minimal separators of $G$.
We can sort the minimal separators with respect to their size in linear time, and therefore start with an edge $ab$ labelled with a max-min separator $S$ and then explore $T_{a}$ and stop either because the whole subtree is labelled with minimal separators contained in $S$, then it suffices to modify the tree,
or because we have found an edge labelled with some edge $xy$ labelled with a minimal separator $S'$ incomparable with $S$. 
In this case, among all edges in $T_{a}$, consider the edge $zt$ labelled with a min-max separator $S''$ incomparable with $S$, and recurse on $zt$.
During this algorithm an edge of $T$ is at most traversed twice, which yields the linearity of the whole process.
\end{proof}

\begin{corollary}
For any chordal graph there exist an elimination scheme that follows a linear extension of the containment ordering
of the minimal separators. It can be computed in $O(n.m)$.
\end{corollary}
\begin{proof}
It is well-known, that one can produce elimination scheme on the following way. Take any maximal  clique tree $T$ of a chordal graph $G$, and let $C$ be a leaf of this tree, attach to the tree via the minimal separator $S$. Successively prune all vertices in $C-S$ and recurse on $T-C$ the maximal clique tree of  $G-\{C-S\}$. To finish the proof it suffices to apply the above theorem. Each time the above algorithm is applied requires $O(n+m)$, this yields the complexity.
\end{proof}

It should be noticed that not every linear extension of the  containment ordering can be obtained with an elimination scheme.

\section{Reversible elimination schemes}

A  reversible elimination scheme is just an ordering of the vertices which is simplicial in both directions.
As shown by the graph  called 3-sun, there exist graphs for which one can prove that there is no
reversible elimination scheme. A vertex is said to be \textbf{bisimplicial} if its neighbourhood can be 
partionned into two cliques.
Furthermore, if a graph $G$ admits such a reversible elimination scheme, this implies that each vertex is either simplicial or bisimplicial. Therefore such a graph cannot contain any claw ($K_{1,3}$) as subgraph.

\begin{figure}[h!]
	\begin{center}
		\subfigure[][Claw: $K_{1,3}$]{\includegraphics[scale=1.0]{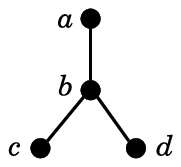}}
\quad
		\subfigure[][3-Sun]{\includegraphics[scale=1.0]{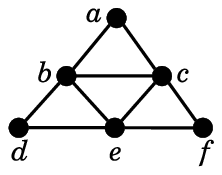}}	
\quad
		\subfigure[][$net$]{\includegraphics[scale=1.0]{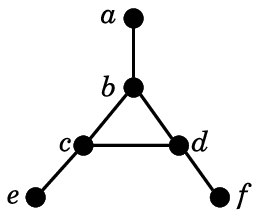}}
	\caption{Forbidden induced subgraphs for proper interval graphs.s}
	\label{fig:ForbiddenProperIntervalGraph}
	\end{center}
\end{figure}

\begin{theorem}
A graph $G$ admits a reversible ordering if and only if $G$ is 
proper interval graph.
\end{theorem}
\begin{proof}
Let us consider a unit interval graph $G$ and one of its unitary interval representation. Therefore to each vertex $x \in G$ we can associate
an interval $I(x)=[left(x), right(x)]$ of length one of the real line, such that $xy $ is an edge iff  $I(x) \cap I(y)\neq \emptyset$.
Let us consider the total ordering $\tau$ of the vertices of $G$ defined as follows:   $x \leq_{\tau}y $  iff  $(right(x) < right(y))$.
Let $x$ be the first vertex of this ordering, clearly its neighborhood is a clique. Thus  $\tau$ is an elimination scheme. Reversibility is straightforward.
Conversely let us  proceed by contradiction. Let us assume that $G$ admits a reversible elimination 
ordering and that $G$ is not a proper interval graph. As proper interval 
graph admit a characterization by forbidden induced subgraphs, we can assume 
that our graph contains one of the graph as a subgraph. The forbidden sugraphs 
for proper interval graphs are the net, the claw and the sun of size $3$. These 
graphs are depicted in figure \ref{fig:ForbiddenProperIntervalGraph}.
So to prove our claim it is sufficient to see that none of these graphs 
admit a reversible elimination ordering.
For the claw, we already noticed it.
Considering the 3-sun, it is easy to check that each vertex 
is bisimplicial. If we consider the sugraph induced by $\{a,b,c,d,e\}$, 
this graph forms the bull. And this graph admit only one 
reversible elimination ordering which is $a,b,d,c,e$. 
To convince ourself $a$ and $e$ has to be the extremities of the 
ordering ($d$ is not a good candidate since it is not simplicial 
in the whole graph). Then to satisfy $b$, since $a$ is already 
positionned $c$ and $d$ have to be on the right. In the same 
way to satisfy $c$, since $e$ is already positionned 
$b$ and $d$ have to be on the left. Finally the only 
ordering to fullfill all the constraints is $a,b,d,c,e$. But now, 
when we want to add $f$, each position in the previous order 
will violate the constraint for at least one vertex. A contradiction.
For the net, the proof is similar. \end{proof}

\paragraph{Acknoledgements:}
We are grateful to B. L\'{e}v\^{e}que for pointing out useful references.

\bibliographystyle{plain} 
% \bibliography{MinMaxsep}
\bibliography{simplicial2}
\end{document}